 \documentclass[final,5p,times,twocolumn]{elsarticle}

\usepackage{graphicx}
\usepackage{epstopdf}
\usepackage{amsmath}
\usepackage{amssymb}
\usepackage{widetext}
\usepackage{xcolor}

\journal{Physics of the Dark Universe}

\begin{document}

\begin{frontmatter}

\title{Rotating black holes and exotic compact objects in the Kerr/CFT correspondence within Rastall gravity}

\author[label0,label1,label2,label3]{Muhammad F. A. R. Sakti\corref{cor1}}
\address[label0]{High Energy Physics Theory Group, Department of Physics, Faculty of Science, Chulalongkorn University, Bangkok 10330, Thailand}
\address[label1]{Department of Physics, Universitas Indonesia, Depok, Indonesia 16424}
\address[label2]{Theoretical Physics Laboratory, THEPi Division, Institut Teknologi Bandung, Jl. Ganesha 10 Bandung, Indonesia 40132}
\address[label3]{Indonesia Center for Theoretical and Mathematical Physics (ICTMP), Institut Teknologi Bandung, Jl. Ganesha 10 Bandung, Indonesia 40132}

\cortext[cor1]{Corresponding author}

\ead{fitrahalfian@sci.ui.ac.id}

\author[label2,label3]{Agus Suroso}
\ead{agussuroso@fi.itb.ac.id}

\author[label1]{Anto Sulaksono}
\ead{anto.sulaksono@sci.ui.ac.id}

\author[label2,label3]{Freddy P. Zen}
\ead{fpzen@fi.itb.ac.id}

\begin{abstract}
Quantum gravitational effects on the near horizon may alter the black hole's horizon drastically to be partially reflective, portrayed by a quantum membrane.  With this modification, the object can be considered as an exotic compact object (ECO). Quantum effects on the strong gravitational regime may also lead to a non-conserved matter tensor that can be described phenomenologically using Rastall gravity. In this work, we study the properties of black holes and ECOs within Rastall gravity using Kerr/CFT correspondence. We systematically investigate the properties of the most general rotating black hole solutions in Rastall gravity, i.e., Kerr-Newman-NUT-Kiselev, and reveal its hidden conformal symmetry. The Cardy microscopic entropy formula and absorption cross-sections from 2D CFT are computed and then matched with gravity calculation. We also extend the dual CFT analysis for studying the properties of ECOs. The existence of the quantum membrane leads to the appearance of the gravitational echoes that is manifested as an oscillatory feature on the absorption cross-section. We compute the absorption cross-section and quasi-normal modes in the dual CFT picture. We also compare the absorption cross-section of ECOs to that of black holes. We find that the Rastall coupling constant plays a significant role for both objects. We also obtain that the echo time delay depends explicitly on the Rastall coupling constant. This coupling constant may play a role to recover the correction on time delay that is believed as a non-linear physics effect. Henceforth, the signature of the Rastall gravity can be probed from the time-delay observation.
\end{abstract}


\end{frontmatter}


\section{Introduction}
\label{sec:intro}
The AdS/CFT correspondence has been discovered as a major advance of a non-perturbative formulation of string theory. This correspondence has given a successful realization of the holographic principle proposed by 't Hooft \cite{tHooft}. It provides a powerful toolkit for investigating strongly coupled theories by using weakly coupled theories or \textit{vice versa}. A more practical application of this correspondence is in the study of black hole's thermodynamics. A very notable breakthrough of this correspondence is the study of Kerr black hole's entropy, known as Kerr/CFT correspondence \cite{Guica2009}. In the Kerr/CFT correspondence of \textcolor{black}{extremal black holes}, it is found that the near-horizon geometry of extremal Kerr black holes \textcolor{black}{(spin $a$ is equal to mass $M$)} exhibits an exact $SL(2,R)\times U(1)$ symmetry leading to the precise computation of Bekenstein-Hawking entropy from Cardy formula from two-dimensional (2D) CFT. This study then has been extended for the Kerr black hole with general \textcolor{black}{value of spin parameter (non-extremal)} \cite{Castro2010}. However, \textcolor{black}{in this non-extremal case,} conformal symmetry cannot be found directly from the geometry of the black holes. The conformal symmetry is stored (hidden) on the solution space of the probe scalar field in the near-horizon region within the low-frequency approximation. The probe scalar field equation can be written in quadratic Casimir operator that satisfies $SL(2,R)\times SL(2,R)$ algebra. The conformal symmetry obtained from this calculation acts locally on the solution space but \textcolor{black}{it} is globally obstructed by the periodic identification of the azimuthal angle $\phi $. This identification leads to the broken  $SL(2,R)\times SL(2,R)$ symmetry to $U(1)\times U(1)$ symmetry by the left- and right-moving temperatures $T_{L,R}$. Using the numerological observation, the central charges of this \textcolor{black}{non-extremal} Kerr black hole are identical to the extremal one. Then the resulting Cardy entropy will exactly match with the Bekenstein-Hawking entropy as the extremal Kerr black holes.
Furthermore, in \cite{HacoJHEP2018}, the authors gave a more enticing derivation of the central charges using the covariant phase space formalism \textcolor{black}{for non-extremal Kerr black holes}. By assuming the Wald-Zoupas counterterms, they managed to compute the left and right central charges. The local conformal symmetry in the Kerr background is also sufficient to obtain the cross-section to that of the gravity computation. For a review regarding this whole Kerr/CFT computation can be referred to Ref. \cite{Compere2017}, while it has also been applied in several black hole solutions in extremal \cite{Hartman2009,Ghezelbash2009,Lu2009,Li2010,Anninos2010,Ghodsi2010,Ghezelbash2012,Astorino2015,Astorino2015a,Siahaan2016,Astorino2016,Sinamuli2016,Sakti2018,SaktiEPJPlus2019,SaktiMicroJPCS2019,SaktiPhysDarkU2020} and generic \cite{ChenLong2010,ChenSun2010,ChenWang2010,WangLiu2010,ChenLongJHEP2010,SetareKamali2010,GhezelbashKamali2010,ChenHuangPRD2010,ChenChen2011,ChenGhezelbash2011,ChenZhangJHEP2011,HuangYuan2011,Shao2011,DeyouChen2011,GhezelbashSiahaan2013,SiahaanAcc2018,Saktideformed2019,SaktiNucPhysB2020} cases.

Intriguingly, the Kerr/CFT correspondence could also be implemented to the object beyond \textcolor{black}{classical} black holes, namely exotic compact object (ECO) as discussed in Ref. \cite{DeyAfshordiPRD2020}. It is believed that the quantum gravitational effects in the near-horizon region may modify the apparent horizon of the \textcolor{black}{classical} black holes leading to potentially observable consequences. Due to the quantum gravitational effects in the near-horizon region, the structure on the event horizon may alter drastically to a \textcolor{black}{reflective quantum membrane (an infinitesimal reflective layer with an arbitrary reflectivity and placed slightly (in the order of Planck length) in front of the would-be horizon of the classical black hole)}. This description is also considered as an alternative way to solve the black hole information paradox. \textcolor{black}{Somehow, this reflective membrane does not correspond to the firewall proposal \cite{AlmheiriJHEP2013}. The firewall in the black hole is a hypothetical phenomenon where an observer falling into a black hole encounters high-energy radiation at the event horizon. This suggests that the entanglement must somehow get broken immediately between the infalling particle and the outgoing particle leading to the release of large amounts of energy and creating a firewall at the black hole event horizon. There are also some proposals addressing the modification of the black hole's structure in order to solve that problem such as gravastars \cite{MazurMottola2004,MazurMottolaCQG2015,SaktiSulaksonoPRD2021}, fuzzballs \cite{MathurForsch2005,MathurTurtonJHEP2014}, 2-2 holes \cite{22holes}, and Kerr-like wormholes \cite{BuenoCanoPRD2018}.} The near-horizon modification that results in a quantum reflective membrane on the ECOs can originate as finite-size effects in CFT living on a torus with finite length $L$ and inverse of the temperature $1/T$, in contrary with the CFT dual of black holes where we need to keep one of torus parameters infinite.
Due to the existence of the quantum membrane, it is expected that the waves traveling to the would-be horizon will get partially reflected. This feature will lead to the presence of gravitational echoes in the postmerger ringdown signal of a binary black hole coalescence \cite{CardosoFranzinPRL2016,OshitaAfshordiPRD2019,CardosoHopperPRD2016,AbediDykaarPRD2017,WangOshitaPRD2020}. Such detection of echoes has made the investigation of quantum gravitational effects of astrophysical objects more promising. This fact can also be a well-turned attempt to seek the unification between gravitational and quantum theories. One may expect that such unification can be probed in the observational signatures of the ECOs. In particular, the realization of the observational signatures of the ECOs lies in the detection of the quasi-normal modes (QNMs) and echo time delay \cite{WangOshitaPRD2020,DeyChakrabortyPRD2020,DeyAfshordiPRD2020,DeyBiswasPRD2021,OshitaAfshordiPRDQuantum2020}. The QNM spectrum primarily dominates the final ringdown phase of a binary black hole coalescence where we need to impose the boundary conditions at the horizon and asymptotic infinity. For the ECO, the presence of the reflective membrane modifies the boundary condition at the horizon for which the ingoing waves can get partially reflected by the membrane and reach back to the asymptotic observer at infinity with a time delay.

Precise detection of the QNM spectrum can be obtained from the analysis of the ringdown spectrum, which may give a adequate idea about the nature of the black holes or ECOs. Moreover, it may provide signatures to test general relativity and beyond general relativity. For example, the calculation of QNMs might be used to probe the extra dimension from the ECOs in the braneworld scenario \cite{DeyBiswasPRD2021,DeyChakrabortyPRD2020}. Besides the QNM spectrum, one can expect the time delay is also sufficiently beneficial for probing the theories beyond general relativity. As noted in \cite{AbediDykaarPRD2017}, the time between the main merger event and the first echo could be influenced by the non-linear physics effect. Furthermore, the influence from non-linear physics may cause the change of magnitude of the echo time delay at 2$\%$-3$\%$. 

This work investigates the black holes and ECOs in such a scenario within Rastall theory of gravity. Rastall gravity considers the generalization of the general relativity, especially in the assumption of the conserved covariant derivative equation of the matter tensor. Originally, Rastall \cite{Rastall1972,Rastall1976} considered that the covariant divergence of matter tensor is not null, which given by
\begin{equation}
\nabla_\mu T^{\mu\nu}=\lambda \nabla^\nu R, 
\end{equation}
where $\lambda $ is a constant that quantifies the deviation from general relativistic theory, and the covariant divergence is proportional to the covariant derivative of Ricci scalar $R$. Later on, we denote $\kappa\lambda $ as Rastall coupling constant where $\kappa=8\pi$. Due to the non-vanishing divergence of matter tensor, the Einstein field equation will be modified as
\begin{equation}
G_{\mu\nu} +\kappa\lambda \textcolor{black}{R} g_{\mu\nu}=\kappa T_{\mu\nu}.
\end{equation}
From above equation, one can find $R=\kappa T/(4\kappa\lambda -1) $. Rastall noted that in this theory, $\kappa\lambda=1/4$ is the exemption because this results in an irregular Ricci scalar. \textcolor{black}{For the case $\kappa\lambda =1/4$, this is explained as unimodular gravitational theory \cite{NgvanDamPRL1990,NgvanDamJMP1991unimodular} and will not be considered within this paper}. Obviously, Rastall gravitational theory considers that the matter tensor is coupled non-minimally with gravity as $f(R,T)$ theory, so Rastall gravity can be considered as a specific condition for $f(R,T)$ theory \cite{FisherCarlsonPRD2019,MoraesSantosGRG2019,ShabaniEPL2020,HarkoMoraesPRD}. In a strong gravitational regime, the covariant derivative of the matter tensor might not be satisfied. Nonetheless, Rastall gravity is still valid in the flat space-time and satisfies the conserved covariant derivative of matter tensor. Hence, this theory is still feasible as the generalization of general relativity. Rastall theory has also been applied to several astrophysical objects \cite{SaktiAnnPhys2020,Abbas2018,Abbas2019,Oliveira2015,Moradpour2017a,MaulanaSulaksonoPRD2019,HadyanIJMPD2020} and cosmology \cite{Moradpour2016,Batista2012,Moradpour2017} in different aspects. Recently, Visser \cite{Visser2018} claimed that Rastall gravity is equivalent with Einstein gravity. However, it was argued by Darabi \textit{et al.} \cite{Corda2018} that Visser's conclusion is not correct by comparing these two gravitational theories. Henceforth, whether these two theories are equivalent or not, Rastall gravity may face the challenges of cosmological or astrophysical observations as the general relativity.

In a gravitational system, quantum effects may give rise to the violation of the conservation of the matter tensor. Rastall gravity can be considered as a phenomenological way to identify the effects of quantum fields in the curved space-time in a covariant formulation, for example, in the case of gravitational anomalies \cite{Bertlemann,Birrell}. Since the quantum gravitational effects may modify the apparent horizon of the black holes giving rise to the presence of a \textcolor{black}{reflective} quantum membrane, we may expect that quantum effects are also manifested as a non-vanishing Rastall coupling constant. Hence, within the framework of black holes and ECOs, the Rastall coupling constant is expected to give a distinctive signature on the observables, such as the QNM spectrum and echo time delay. 

In the following, we address the implementation of Kerr/CFT correspondence to investigate different astrophysical compact objects.
Firstly, we investigate the hidden conformal symmetry of rotating black holes in Rastall gravity. We choose the Ker-Newman-NUT-Kiselev (KNUTK) black hole solutions in Rastall gravity obtained in \cite{SaktiAnnPhys2020}. These solutions portray the charged rotating and twisting black holes environed by the quintessential field with a non-vanishing Rastall coupling constant. The entropy and absorption cross-section of these generic black hole solutions is computed. This computation denotes that these black hole solutions are holographically dual with 2D CFT. Secondly, since the black hole solutions can be described in a 2D CFT picture, we apply a similar duality calculation to the ECO-type of KNUTK black hole solutions in Rastall gravity where there is a reflective quantum membrane close to the would-be horizon. We calculate the QNM spectrum, absorption cross-section, and echo time delay in the 2D CFT picture. Generally, it is found that the Rastall coupling constant plays a significant role on the observables along with the other black hole parameters.

The organization of this paper is given as follows. In the next section, we focus our investigation on KNUTK black hole solutions in Rastall gravity. This section is to reveal the hidden conformal symmetry on the wave equation. We also attest the holographic dual of the black holes by computing the microscopic entropy and absorption cross-section. The subsequent section is focused on the study of the KNUTK-like ECO in Rastall gravity. We perform the duality calculation to compute the QNMs. Then the absorption cross-section is also investigated to see the effects of the quantum membrane. Afterward, we also delve into the echo time delay of this ECO. Eventually, we summarize the whole paper in the last section.

\section{Rotating black holes in Rastall gravity}\label{gen-hidden}

The KNUTK black holes are the exact solution to Einstein-Maxwell-perfect fluid equations with non-vanishing NUT charge in Rastall gravity for a non-vanishing Rastall coupling constant. These black hole solutions are obtained in \cite{SaktiAnnPhys2020} by using Demia{\'n}ski-Newman-Janis algorithm to the Reissner-Nordstr{\"o}m-Kiselev black holes in Rastall gravity. The solutions portray the space-time for a rotating electrically and magnetically charged massive object with a twisting NUT charge surrounded by a quintessential field. Kiselev \ first derived the quintessential field in the black hole solutions \cite{Kiselev2003}, so that it is common to say the quintessential black holes as Kiselev black holes. The line element of KNUTK \textcolor{black}{blak hole} space-time in Rastall gravity reads as \cite{SaktiAnnPhys2020}
\begin{eqnarray}
ds^2 &=& - \frac{\Delta}{\varrho^2} \left[ dt \!-\! \{a \sin^2\theta + 2n(1-\cos\theta) \} d\phi \right]^2 +  \frac{\varrho ^2}{\Delta}dr^2 \nonumber\\ 
& &+ \varrho ^2 d\theta^2  + \frac{\sin^2\theta}{\varrho ^2}\left[adt - \{r^2+(a+n)^2 \} \right] d\phi^2,\label{KNNUTAdSmetric} \
\end{eqnarray}
where
\begin{equation}
\varrho^2=r^2+(n+a\cos\theta)^2, ~~ \Delta =r^2- 2M r+\textcolor{black}{a^2} +e^2 +g^2 -n^2 -\alpha r^{\upsilon}, \nonumber\
\end{equation}
\begin{equation}
\upsilon = \frac{1-3\omega_q}{1-3\kappa\lambda (1+\omega_q)}.\label{rho} \nonumber\
\end{equation}
The parameters $a$ and $n$ represent the rotation parameter and  NUT charge or twisting parameter. The mass parameter, electric charge, and magnetic charge of space-time are given by $M$, $e$, and $g$, respectively. $ \alpha $ is the intensity of the quintessential field, and $ \omega_q $ is the equation of the state's parameter of the quintessential field. The quintessential field in the black hole solutions is believed to be the hypothetical realization of the dark energy represented by the scalar field in the local astrophysical objects for the given equation of state \cite{Kiselev2003}. \textcolor{black}{It is worth pointing out that the space-time metric (\ref{KNNUTAdSmetric}) is not asymptotically flat due to the presence of NUT charge and for arbitrary value of $\omega_q$ and $\kappa\lambda$.}

The space-time metric (\ref{KNNUTAdSmetric}), in the limit of $a=0$ reduces to the Reissner-Nordstrom-NUT-Kiselev black hole in Rastall gravity. Moreover, it reduces to Taub-NUT-Kiselev space-time and Schwarzschild-Kiselev space-time in Rastall gravity, in the limits of $a=0,~e=0,~g=0$  and $a=0,~e=0,~g=0,~n=0$, respectively. It is clearly seen that for $a=0,~e=0,~g=0,~n=0,\alpha =0$, it reduces to Schwarzschild space-time in Rastall gravity. \textcolor{black}{When the quintessential field is absent in metric (\ref{KNNUTAdSmetric}), the parameter $\omega_q$ vanishes and we can treat $\alpha$ just as an integration constant. Basically, $\alpha$ appears as an integration constant \cite{Kiselev2003} and it appears too when considering only the Rastall gravity. This integration constant can be considered as cosmological constant argued in Ref. \cite{Visser2018}. However, this is still debatable as you can see in Ref. [78] that the authors contrarily argued to the opinion in Ref. \cite{Corda2018}. Hence, our analysis within this paper will try to provide the possibility of probing Rastall gravity in the black hole and ECO.}

The electromagnetic potential \textcolor{black}{related to the space-time metric} (\ref{KNNUTAdSmetric}) is given by \cite{SaktiAnnPhys2020}
\begin{eqnarray}
A_\mu dx^\mu &=& \frac{-er\left[a dt- \left( (a+n)^2 - (n+a \text{cos}\theta)^2 \right) d\phi \right]}{a\varrho ^2}  \nonumber\\
& &- \frac{g(n+a\text{cos}\theta)\left[a dt- \left( r^2 + (a+n)^2 \right) d\phi \right]}{a\varrho ^2} .\ \label{eq:electromagneticpotential}
\end{eqnarray}
The Bekenstein-Hawking entropy, Hawking temperature, and angular velocity on the event horizon ($r_+$) of the \textcolor{black}{classical KNUTK} black hole,
are given by 
\begin{eqnarray} \label{en}
&&S_{BH}=\frac{A_{BH}}{4}=\pi \left[r_+^2+(a+n)^2\right],\label{en1}\\
&&T_H=\frac{2(r_+-M)-\alpha \upsilon r_+^{\upsilon -1}}{4\pi[r_+^2+(a+n)^2]},\label{en2}\\
&&\Omega_H=\frac{a}{r_+^2+(a+n)^2},\label{en3}\
\end{eqnarray} 
respectively. \par
\subsection{Hidden conformal symmetry}
Unlike the extremal black holes, the conformal symmetries are hidden on the scalar wave equation of the black hole's background. To investigate and explore the hidden conformal symmetry, we consider a massless neutral scalar probe in the background of the KNUTK black holes in Rastall gravity (\ref{KNNUTAdSmetric}). Note that one can also assume a (electrically and magnetically) charged scalar probe to explore the conformal symmetries in different pictures. Nonetheless, within this paper, we focus our analysis only for the angular momentum picture. The massless scalar field equation for the scalar probe, is given by
\begin{equation}
\nabla_{\alpha} \nabla^{\alpha}\Phi = 0\label{KG1}.
\end{equation}
We notice the KNUTK black hole (\ref{KNNUTAdSmetric}) have two translational Killing vectors i.e., $\partial_t$ and $\partial_\phi$. Hence, we can separate the coordinates in the scalar field as 
\begin{equation}
\Phi(t, r, \theta, \phi) = \mathrm{e}^{- i \omega t + i m \phi} R(r) S(\theta)\label{phi-expand1}.
\end{equation}
Plugging Eq. (\ref{phi-expand1}) into Eq. (\ref{KG1}), leads to two differential equations i.e., the angular $S(\theta)$ and radial $R(r)$ wave functions,
\begin{widetext}
	\begin{equation}
	\frac{1}{\sin\theta} \partial_\theta (\sin\theta \partial_\theta )S(\theta) - \Bigg[ \frac{m^2}{\sin^2\theta}+\frac{\left[2n(\cos\theta-1)-a \sin^2\theta\right]^2\omega^2}{\sin^2\theta} \Bigg] S(\theta) - \left[ 2am\omega \cos\theta (4n+a\cos\theta) + K_{l} \right]S(\theta) =0  \label{angular1}, \end{equation}
\end{widetext}
	\begin{equation} \partial_r (\Delta \partial_r)R(r) + \bigg[ \frac{[ (r^2 + (a+n)^2) \omega - m a ]^2}{\Delta} + 2 a m \omega - K_{l} \bigg] R(r) = 0\label{radial1}, \end{equation}
where $K_{l}$ is the separation constant. 

The hidden conformal symmetries can only be revealed when we look at the near-horizon region of the scalar wave equation. Hence, we consider the radial equation (\ref{angular1}) at the near-horizon region, which is defined by $ \omega \textcolor{black}{r_+} \ll 1 $. Moreover, we consider the low-frequency limit for the scalar field $ \omega M \ll 1 $. Consequently, we also impose $ \omega a \ll 1, \, \omega e \ll 1,  \omega g \ll 1,  \omega \sqrt{\alpha} \ll 1, $ and $ \omega n \ll 1 $. As the study of the hidden conformal symmetries on Kerr-AdS black hole's family, the function $ \Delta  $ should be approximated into two horizons. It is necessary to approximate $\Delta $ by a quadratic polynomial of $ (r-r_+) $ to set up the radial equation in a suitable form for exploring its hidden conformal symmetry. We can 
approximate $\Delta $ by a quadratic polynomial in  $( r-r_+) $ as,
\begin{equation}
\Delta \simeq K(r-r_+)(r-r_*), \label{APP}
\end{equation} 
where 
\begin{equation} 
K = 1- \frac{\alpha \upsilon (\upsilon-1)r_+^{\upsilon-2}}{2}, \nonumber\
\end{equation}
\begin{equation} 
r_* = r_+ - \frac{1}{K r_+}\left[r_+^2 - a^2 - e^2 -g^2 +n^2+\alpha(1-\upsilon)r_+^\upsilon  \right]. \nonumber\
\end{equation}
It is worth noting that $ r_* $ is not the inner horizon. In terms of $ r_+ $ and $ r_* $, we can write the Hawking temperature as
\begin{equation}
T_H = \frac{K(r_+ -r_*)}{4\pi [r_+^2 + (a+n)^2]}.
\end{equation}
Such an approximation was used before for the quartic metric function of the Kerr-Newman-AdS \cite{ChenLongJHEP2010,ChenChen2011}, Kerr-NUT-AdS \cite{ChenGhezelbash2011}, and Kerr-Newman-NUT-AdS \cite{Saktideformed2019,SaktiNucPhysB2020} black holes. Using (\ref{APP}), the radial equation (\ref{radial1}) reduces to 
\begin{equation}
\left[ \partial_r \left[(r-r_+)(r-r_*) \partial_r\right] + \frac{r_+ - r_*}{r - r_+} A+\frac{r_+ - r_*}{r - r_*} B + C \right]  R(r) = 0 , \label{radeq} \
\end{equation} 
where
\begin{equation}
A= \frac{ \left[ (r_+^2 \!+\! (a+n)^2) \omega - a m \right]^2}{K^2 (r_+ - r_*)^2}, \nonumber\
\end{equation}
\begin{equation}
B= \frac{\left[ (r_*^2 \!+\! (a+n)^2) \omega - a m \right]^2}{K^2(r_+ - r_*)^2}, ~~~ C= \frac{-K_l}{K} . \label{C}\nonumber\
\end{equation}
We should note that not all generic rotating black holes are holographically dual to two different CFTs.  Specifically, the four-dimensional charged rotating magnetized Kerr black holes cannot be written in the above form to reveal the hidden conformal symmetries \cite{Siahaan2016}. It is important to note also that we do not explore the angular wave equation because this does not have $ SL(2,R)\times SL(2,R) $ isometry, yet $ SU(2)\times SU(2) $ isometry \cite{LoweSkanataPRD2014}.

In order to reveal the hidden symmetry, we perform the following coordinate transformations for the generic black holes \cite{Castro2010,Compere2017}
\begin{eqnarray}
&& \omega^+ = \sqrt{\frac{r-r_+}{r-r_*}}e^{2\pi T_R \phi + 2 n_R t}, \label{con1}\\
&& \omega^- = \sqrt{\frac{r-r_+}{r-r_*}}e^{2\pi T_L \phi + 2 n_L t}, \label{con2}\\
&& y = \sqrt{\frac{r_+ -r_*}{r-r_*}}e^{\pi (T_L +T_R) \phi + (n_L + n_R) t}. \label{conformalcoord}\
\end{eqnarray}
In terms of the new conformal coordinates $\omega^+,\,\omega^-$ and $y$, we may define three locally conformal operators 
\begin{eqnarray}
&& H_1 = i \partial_+, \label{vec1}\\
&& H_{-1} = i \left(\omega^{+2}\partial_+ + \omega^{+}y\partial_y - y^2 \partial_- \right), \label{vec2}\\
&& H_0 = i \left(\omega^{+}\partial_+ + \frac{1}{2}y\partial_y \right), \label{vec3}
\end{eqnarray}
as well as 
\begin{eqnarray}
&& \bar{H}_1 = i \partial_-, \label{vvec1}\\
&& \bar{H}_{-1} = i \left(\omega^{-2}\partial_- + \omega^{-}y\partial_y - y^2 \partial_+ \right), \label{vvec2}\\
&& \bar{H}_0 = i\left(\omega^{-}\partial_- + \frac{1}{2}y\partial_y \right), \label{vvec3}
\end{eqnarray}
\textcolor{black}{where  $\partial_+ = \partial/\partial \omega^+$ and $\partial_- = \partial/\partial \omega^-$.} The set of operators (\ref{vec1})-(\ref{vec3}) satisfy the $ SL(2,R) $ Lie algebra
\begin{eqnarray}
\left[H_0,H_{\pm 1} \right] = \mp iH_{\pm 1}, ~~~ \left[H_{-1},H_1 \right]=-2iH_0, 
\end{eqnarray}
while a similar $ SL(2,R) $ algebra exists for the set of operators (\ref{vvec1})-(\ref{vvec3}). From any of two sets of operators, we can obtain the quadratic Casimir operator that reads as
\begin{eqnarray}
\mathcal{H}^2 &=& \bar{\mathcal{H}}^2 = - H_0^2 + \frac{1}{2}(H_1 H_{-1} + H_{-1} H_{1}) \nonumber\\
&=& \frac{1}{4}(y^2 \partial_y^2 - y\partial_y)+y^2\partial_+ \partial_-. \label{quadraticCasimir}
\end{eqnarray}
It will be easier to notice the relation between the quadratic Casimir operator (\ref{quadraticCasimir}) and the radial Eq. (\ref{radeq}) in old coordinates $ (t,r,\phi) $. In old coordinates, the quadratic Casimir operator is given by
\begin{eqnarray}
\mathcal{H}^2 &=& (r-r_+)(r-r_*) \partial_r^2 + (2r -r_+ -r_*)\partial_r \nonumber\\
& + & \frac{r_+ -r_*}{r-r_*}\left(\frac{n_L-n_R}{4\pi G}\partial_\phi -\frac{T_L-T_R}{4G}\partial_t \right)^2 \nonumber\\
& -& \frac{r_+ -r_*}{r-r_+}\left(\frac{n_L+n_R}{4\pi G}\partial_\phi -\frac{T_L+T_R}{4G}\partial_t \right)^2 , \label{CAS}
\end{eqnarray}
where $ G = n_L T_R - n_R T_L $ and $ n_L, T_R, n_R, T_L $ are constants.

We have found that the radial Eq. (\ref{radeq}) could be re-written in terms of the $ SL(2,R) $ quadratic Casimir operator as $ \mathcal{H}^2 R(r)=\bar{\mathcal{H}}^2 R(r)= -C R(r) $ where we should identify the constants as
\begin{equation}
n_L = -\frac{K}{2(r_+ + r_*)},~~~ n_R =0, \label{nJ}
\end{equation}
\begin{equation}
T_L = \frac{K[r_+^2 + r_*^2+2(a+n)^2]}{4\pi a(r_+ + r_*)},~~~ T_R =\frac{K(r_+ - r_*)}{4\pi a}. \label{eq:tempCFTKNNUT}
\end{equation}
We may identify $ T_{R,L} $ as the CFT temperatures that arise as a result of the spontaneously symmetry breaking of the partition function on $ SL(2,R)\times SL(2,R) $ theory to the partition function of $ U(1)\times U(1) $ CFT. By periodic identification of the azimuthal coordinate $ \phi \sim \phi +2\pi $, the $ SL(2,R)\times SL(2,R) $ symmetry breaks down to $ U(1)\times U(1) $ symmetry.


\subsection{Microscopic entropy description}\label{entropy}

In the previous subsection, we have shown the hidden conformal symmetry where the dual CFTs could describe the KNUTK black holes in Rastall gravity. For the extremal rotating black holes in Rastall gravity \cite{SaktiPhysDarkU2020}, the CFTs have only a non-zero left temperature. Nevertheless, the dual CFTs have finite and non-zero left- and right-moving temperatures for the generic black holes.  Another essential piece of information for any CFT is the central charge of the underlying theory. For the extremal black holes, the central charges of the dual CFT could be derived by using analysis of the asymptotic symmetry group \cite{Guica2009,Compere2017}. 

In order to use the analysis, we need to work on the near-horizon region.  To find the near-horizon geometry of the extremal KNUTK black holes in Rastall gravity (\ref{KNNUTAdSmetric}),  we employ the following coordinate transformations
\begin{eqnarray}
r = r_+ +\epsilon r_0 y,~~ t = \frac{r_0}{\epsilon}\tau, ~~ \phi = \varphi + \frac{\Omega_H r_0}{\epsilon}\tau, \label{extremaltransformation} \
\end{eqnarray}
where $r_0^2={r_+^2+(a+n)^2}$ and $ \epsilon \rightarrow 0 $ denotes the near-horizon limit.
In the near-horizon coordinates, the metric (\ref{KNNUTAdSmetric}) becomes
\begin{equation}
ds^2 = \Gamma(\theta)\left(-y^2 d\tau^2 + \frac{dy^2}{y^2} + K d\theta ^2 \right) +\gamma(\theta) \left(d\varphi +p y d\tau\right)^2, \label{extremalmetric}\
\end{equation}
where
\begin{equation}
\Gamma(\theta)=\frac{\varrho_+ ^2}{K},  ~~~\gamma(\theta)=\frac{r_0^4  \sin^2\theta}{\varrho_+^2}, \nonumber\
\end{equation}
\begin{equation}
\varrho_+^2 = r_+^2 + (n + a\cos\theta)^2, ~~~p = \frac{2ar_+}{K r_0^2}, \label{GAM1}\nonumber\
\end{equation}
and we have used the scaling $ d\tau \rightarrow d\tau/K $. In the near-horizon limit, the gauge field (\ref{eq:electromagneticpotential}) is given by
\begin{equation}
A_\mu dx^\mu = f(\theta) \left(d\varphi +\hat{p}yd\tau \right)-\frac{e[r_+^2-(a+n)^2]}{[r_+^2+(a+n)^2]}d\varphi ,\ \label{eq:nearhorizonelectromagneticpot1}
\end{equation}
where
\begin{eqnarray}
f(\theta) =\frac{er_0^2\left[(r_+^2-n^2)-a(2n+a\text{cos}\,\theta)\text{cos}\,\theta \right]}{2ar_+ \rho_+ ^2}. \label{eq:ftheta} \
\end{eqnarray}
As denoted in \cite{Hartman2009}, the central charge can be calculated directly from the near-horizon geometry of the extremal black holes. Hence, the central charge of the CFT is given by
\begin{equation}
c_L = 3p\int^\pi_0 d\theta\sqrt{\Gamma(\theta) \gamma(\theta)K} = \frac{12ar_+}{K}. \label{c-calc}
\end{equation}

For generic rotating black holes, CFTs are represented by left- and right-moving central charges. So, from (\ref{c-calc}), we can propose the following left- and right-moving central charges for the CFT dual to the generic black holes as
\begin{equation}
c_L = c_R = \frac{6a(r_+ +r_*)}{K}. \label{centralchargeJpic}\
\end{equation}
Note that there is a restriction on the central charge in order to be finite in which $ K \neq 0 $ or $ \alpha \upsilon (\upsilon-1)r_+^{\upsilon-2} \neq 2 $. So, besides $ \kappa\lambda=1/4 $ \cite{Rastall1972,Rastall1976}, we can also obtain the bound of Rastall coupling constant and $ \omega_q $ in this correspondence.

Another way to compute the central charges given in \cite{HacoJHEP2018} where they extend to calculate the Wald-Zoupas counterterms within the covariant formalism that leads to the well-defined central charges.  Our proposal for the central charges (\ref{centralchargeJpic}) together with the CFT temperatures (\ref{eq:tempCFTKNNUT}) 
lead to the Cardy entropy for the dual CFT to the KNUTK black holes in Rastall gravity, as 
\begin{equation}
S_{CFT}=\frac{\pi^2}{3}(c_L T_L +c_R  T_R )=\pi [r_+^2+(a+n)^2].
\end{equation}
It is in a perfect agreement with the Bekenstein-Hawking entropy.

\subsection{Absorption cross-section}\label{scat}
Another realization of the correspondence between rotating black holes in Rastall gravity and the CFTs is the equivalence of the absorption cross-section. We consider the absorption cross-section of scalar probes in the background of the generic black holes. In investigating the absorption cross-section, we consider the near-horizon and asymptotic regions. In the near-horizon region, the approximation of $ \Delta $  in the radial Eq. (\ref{radeq}) is still acceptable. Nevertheless, in the asymptotic region, this approximation breaks down, except for certain conditions. Accordingly, we cannot exploit the usual radial equation to discuss the scattering issue in general. However, for near-extremal black holes, we may pose a well-defined scattering problem. So, in the calculation of the absorption cross-section, we focus on the near-extremal black holes \cite{ChenLongJHEP2010,ChenChen2011,ChenGhezelbash2011,Saktideformed2019,SaktiNucPhysB2020}. In order to do so, we consider the following near-extremal coordinate transformations, as 
\begin{equation}
r = \frac{r_+ + r_*}{2}+\epsilon r_0 y, ~ r_+ -r_* = \mu \epsilon r_0, ~ t = \frac{r_0}{\epsilon}\tau, ~ \phi = \varphi + \frac{\Omega_H r_0 }{\epsilon}\tau, \label{nearextremaltransformation} \
\end{equation}
where $\mu$ is the parameter which quantifies the extremality. We also consider the scalar probe with frequencies around  the super-radiant bound 
\begin{equation}
\omega = m\Omega_H +\hat{\omega}\frac{\epsilon}{r_0},
\end{equation}
where $\Omega_H$ are given by (\ref{en3}). We can re-write the radial Eq. (\ref{radeq}) by
\begin{equation}
\left[\partial_y\left(y-\frac{\mu}{2}\right)\left(y+\frac{\mu}{2}\right)\partial_y +\frac{A_s}{y-\frac{\mu}{2}}+\frac{B_s}{y+\frac{\mu}{2}}           +C_s\right]R(y) =0,\label{radialequationnearext}
\end{equation}
where
\begin{eqnarray}
A_s = \frac{\hat{\omega}^2}{\mu}, ~~~ B_s = -\mu \left(\frac{\hat{\omega}}{\mu}-\frac{2 m\Omega_H  r_+}{K} \right)^2, \nonumber\
\end{eqnarray}
and $ C_s $ is the new separation constant. We also use the coordinate transformation $ z =(y-\mu/2)/(y+\mu/2) $. Now the radial Eq. (\ref{radialequationnearext}) becomes
\begin{equation}
\left[z(1-z)\partial_z^2 + (1-z)\partial_z +\frac{\hat{A_s}}{z}+\hat{B_s}+\frac{C_s}{1-z}\right]R(z) =0,\label{radialequationnearext0}
\end{equation}
where
\begin{eqnarray}
\hat{A_s} = \frac{\hat{\omega}^2}{\mu ^2}, ~~~\hat{B_s} = -\left(\frac{\hat{\omega}}{\mu}-\frac{2 m\Omega_H r_+}{K} \right)^2. \nonumber\
\end{eqnarray}
The differential Eq. (\ref{radialequationnearext0}) possesses the solutions as given by
\begin{equation}
R(z)= z^{-i\alpha_s}(1-z)^{\beta_s}F(a_s,b_s,c_s;z), \label{sol}
\end{equation}
where $ F(a_s,b_s,c_s;z) $ is the hypergeometric function where
\begin{equation}
a_s = \beta_s +i(\gamma_s-\alpha_s), ~b_s =\beta_s -i(\gamma_s + \alpha_s),~c_s = 1- 2i\alpha_s, \nonumber\
\end{equation}
and
\begin{equation}
\alpha_s = \sqrt{\hat{A_s}}, ~\beta_s = \frac{1}{2}\left(1-\sqrt{1-4C_s}\right),~\gamma_s = \sqrt{-\hat{B}_s}. \nonumber\
\end{equation}

In the asymptotic infinity of the radial coordinate $r$ ($y\gg \mu/2 $), the coordinate $z$ becomes 1 that leads to the solutions
\begin{eqnarray}
R(y) \sim D_1 y^{-\beta_s}+D_2 y^{\beta_s -1},
\end{eqnarray}
where
\begin{equation}
D_1 = \frac{\Gamma(c_s)\Gamma(2h-1)}{\Gamma(c_s-a_s)\Gamma(c_s-b_S)}, ~~~ D_2 = \frac{\Gamma(c_s)\Gamma(1-2h)}{\Gamma(a_s)\Gamma(b_s)}. ~~~  
\end{equation}
The conformal weight is given by
\begin{equation}
h= 1-\beta_s=\frac{1}{2}\left(1+\sqrt{1-4C_s}\right).\label{hh}
\end{equation}
Hence, the absorption cross-section of the scalar fields is given by
\begin{equation}
P^{\text{BH}}_{\text{abs}} \sim \left| D_1 \right|^{-2} = \frac{\sinh \left( {2\pi\alpha_s } \right)}{2\pi \alpha_s}\frac{{\left| {\Gamma \left(c_s - a_s  \right)} \right|^2 \left| {\Gamma \left(c_s - b_s \right)} \right|^2 }}{{ | {\Gamma \left( {2h-1} \right)} |^2 }}\label{PabsSL2Z}.
\end{equation}

In \cite{ChenChuJHEP2010}, it is shown that the real-time correlator could be computed holographically
in the bulk. So, we can also find the real-time correlation function $ G_R $ for our black hole solutions. By considering $ D_1 $ as the source and $ D_2 $ as the response, we can obtain
\begin{eqnarray}
G_R \sim \frac{D_2}{D_1} = \frac{\Gamma(1-2h)}{\Gamma(2h-1)} \frac{\Gamma(c_s - a_s)\Gamma(c_s - b_s)}{\Gamma(a_s)\Gamma(b_s)}.
\end{eqnarray}
The absorption cross-section can also be obtained from the correlation function where $P^{\text{BH}}_{\text{abs}}  \sim \text{Im}(G_R)$, the  imaginary part of the correlation function. To further support the correspondence between the rotating black hole (\ref{KNNUTAdSmetric}) and 2D CFT, we show that the absorption cross-section for the scalar fields  (\ref{PabsSL2Z}) can be obtained from the absorption cross-section in a 2D dual CFT. In a 2D CFT with conformal weights $h_{L,R}$ and temperatures $T_{L,R}$, the absorption cross-section for the scalar fields with frequencies $\tilde{\omega}_{L,R}$, is given by \cite{Castro2010,ChenChuJHEP2010}
\begin{eqnarray}
P^{\text{BH}}_{\text{abs}} &\sim &{T _L}^{2h_L - 1} {T _R}^{2h_R - 1} \sinh \left( {\frac{{{\tilde{\omega}} _L }}{{2{T _L} }} + \frac{{{\tilde{\omega}} _R }}{{2{T _R} }}} \right)\nonumber\\
&\times & \left| {\Gamma \left( {h_L + i\frac{{{\tilde{\omega}} _L }}{{2\pi {T _L} }}} \right)} \right|^2 \left| {\Gamma \left( {h_R + i\frac{{{\tilde{\omega}} _R }}{{2\pi {T _R} }}} \right)} \right|^2. \label{Pabs2CFTSL2Z}
\end{eqnarray}

The matching between absorption cross-sections (\ref{PabsSL2Z}) and (\ref{Pabs2CFTSL2Z}) can be obtained by choosing proper left and right frequencies ${\tilde{\omega}} _L,{\tilde{\omega}} _R$. In order to do so, we consider the first law of black hole's thermodynamics which is given by
\begin{equation} \label{BHthermoLaw}
T_H \delta S_{BH} = \delta M - \Omega _H \delta J.
\end{equation}
By varying Cardy entropy, we obtain
\begin{equation} \label{delSCFT}
\delta S_{CFT} = \frac{{\delta E_L }}{{T_L }} + \frac{{\delta E_R }}{{T_R }}.
\end{equation}
We can identify $\delta M$ as $\omega$ and $\delta J$ as $m$ which leads to the identification of $\delta E _{R,L}$ as $\tilde{\omega} _{R,L}$. From the
variations of entropy in (\ref{BHthermoLaw}) and (\ref{delSCFT}), we can find the left and right frequencies as
\begin{eqnarray}
\tilde{\omega} _{L} = \frac{r_+^2 + r_*^2 +2(a+n)^2}{2a}\omega, ~~~ \tilde{\omega} _{R} = \tilde{\omega} _{L}-m.  \label{eq:omegageneral} \
\end{eqnarray}
We note that the conformal weights $ h_{L,R} $ in (\ref{Pabs2CFTSL2Z}) are equal to $h $, as in (\ref{hh}). So, we have found the holographic dual of the KNUTK black holes in Rastall gravity.

\section{Exotic compact object in Rastall gravity}\label{sec:ECO}
In this section, \textcolor{black}{we assume a model of an exotic compact object (ECO) where the exterior space-time metric of this ECO is given by Eq. (\ref{KNNUTAdSmetric}). The near-horizon geometry of the classical black hole given by Eq. (\ref{KNNUTAdSmetric}) is modified due to the existence of quantum membrane structure originating from near-horizon quantum gravitational effects. These quantum gravitational effects lead to the formation of ECO.} This ECO does not possess an event horizon, but a partially reflective membrane located slightly outside the common position of the event horizon. Due to quantum gravitational effects, the reflective membrane can be grasped as an effective description of the quantum corrections on the horizon. This work is inspired by the paper \cite{DeyAfshordiPRD2020} where they studied the Kerr-like ECO. Furthermore, in Rastall theory of gravity, quantum gravitational effects can be manifested as non-conserved divergence of matter tensor, giving rise phenomenologically to non-vanishing Rastall coupling constant.
\begin{figure}[!b]
	\centering
	\includegraphics[scale=0.4]{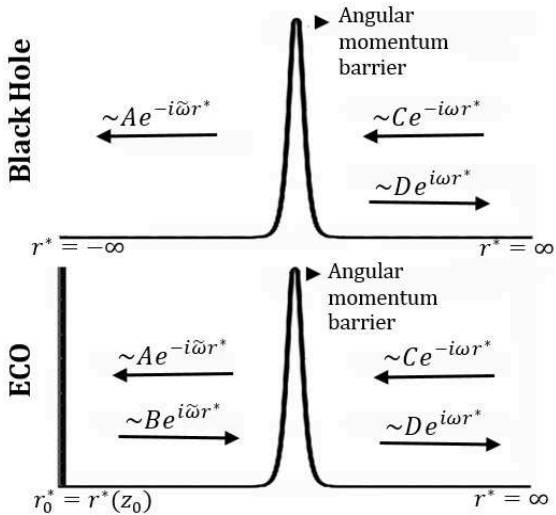}
	\caption{Schematic diagram of the near-horizon region from a black hole and an ECO.}
	\label{fig:BHvsECO}
\end{figure}
The reflective membrane is assumed to be located at \textcolor{black}{ $\bar{r}_0$ where $ \bar{r}_0=r_+ +\delta r_+ $} with reflectivity $ \mathcal{R} $. The constant $ \delta r_+ $ is very small and in the order of Planck length. Due to quantum corrections, the apparent horizon gets shifted as given by \cite{DeyChakrabortyPRD2020,DeyBiswasPRD2021}
\begin{equation}
\delta r_+ \sim \frac{N^2 l_p^2}{M}, \label{epsilondef}\
\end{equation}
where $ N $ is the degree of freedom of the dual CFT living on the torus and $ l_p $ is the Planck length. $ N $ corresponds to length of the torus $ L $ as $ N^2 \sim L^2/l^2_p $. We can calculate $ \delta r_+ $ from
\begin{equation}
\int^{r_+ + \delta r_+}_{r_+} \sqrt{g_{rr}}|_{\theta=0}~ dr \sim \eta l_p , \label{eq:epsilon}\
\end{equation}
where $ \eta $ is just a constant proportional to CFT degree of freedom.

In the black hole space-time, the low- (high-)frequency modes scalar wave will be reflected (transmitted) by the angular momentum barrier, while the higher (lower) frequencies modes will cross the barrier and go to infinity (see Fig. \ref{fig:BHvsECO}). However, due to a quantum membrane for ECO, the low-frequency modes will be trapped between the membrane and the angular momentum barrier. The trapped waves will slowly leak out as repeating echoes. For a scalar perturbation in this ECO background, the asymptotic behavior of the waveform would be given as
\begin{equation}
R \sim \left\{\begin{array}{lr}
\frac{C}{r}e^{-i\omega r^*}+ \frac{D}{r}e^{i\omega r^*}~~~ \text{for}~~~ r^*\rightarrow \infty,\\
Ae^{-i\tilde{\omega}r^*}+ Be^{i\tilde{\omega}r^*}~~~ \text{for}~~~ r^*\rightarrow -\infty,
\end{array}\right.
\end{equation}
where $ \tilde{\omega} $ is the horizon-frame frequency, $ r^* $ is the tortoise coordinate and $ A,B,C,D $ are the asymptotic amplitudes as shown in Fig. \ref{fig:BHvsECO}. The quantum membrane possesses a reflectivity which can be defined in terms of the given asymptotic amplitudes as
\begin{equation}
\mathcal{R}e^{i\pi \delta} \equiv \frac{B}{A}z_0^{2i\sigma}, \label{eq:reflectivity}
\end{equation}
where $ \delta $ is a phase identified by the quantum properties of the ECO, $ z_0 $ is the position of the quantum membrane in the coordinate $ z $, and $ \sigma=(\omega-m\Omega_H)/4\pi T_H $.

One can see the existence of the echoes coming from the massless scalar waves trapped between the quantum membrane and the angular momentum barrier from the calculation of the absorption cross-section. Since the quantum membrane will reflect the massless scalar waves, the absorption cross-section of this ECO will differ from that of a black hole (\ref{PabsSL2Z}), resulting in echoes. As given in \cite{DeyAfshordiPRD2020}, we can write the relation between absorption cross-section of the black holes and ECOs as
\begin{equation}
P^{\text{ECO}}_{\text{abs}} =P^{\text{BH}}_{\text{abs}} \frac{1-|\mathcal{R}|^2}{|1- \mathcal{R}e^{-2ir_0^*(\omega -m\Omega_H)+i\delta }|^2}, \label{eq:PabsECOgen}\
\end{equation}
where $ r_0^* $ is the position of the membrane in terms of tortoise coordinate. It is obvious that for $ \mathcal{R}=0 $, it will reduce to that of classical black holes. It has been derived in \cite{WangOshitaPRD2020,OshitaWangJCAP2020} that the reflectivity for a quantum black hole (ECO) is given by
\begin{equation}
\mathcal{R}\sim e^{-\frac{|\omega -m\Omega_H|}{2T_{QH}}}. \label{eq:reflectivity}\
\end{equation}

$\mathcal{R}$ depends on the frequency and angular momentum of the scalar waves. This finding is the standard Boltzmann reflectivity. Note that $ T_{QH} $ is identified as the quantum horizon temperature and comparable to the Hawking temperature for black holes $ (T_{QH}=\gamma T_{H}) $. As noted in \cite{OshitaAfshordiPRD2019,OshitaAfshordiPRDQuantum2020}, $ \gamma $ is the proportionality constant depending on the dispersion and dissipation effects in graviton propagation.

\subsection{QNMs}\label{sec:QNM}

Gravitational radiation from ECO oscillation exhibits a specific characteristic spectrum independent of the processes that give rise to these oscillations. The QNM spectrum is directly related to the parameters contained by the ECOs. Hence, QNM may carry important information about the near-horizon quantum structure of the ECOs, which is dependent on those parameters. In investigating the black holes, especially using AdS/CFT duality, QNM can be obtained from the poles of the retarded correlation function from the CFTs. In this duality calculation, the quantum gravity in the near-horizon region of the rotating black holes is dual to the 2D CFT. It has been shown in \textcolor{black}{ \cite{BirminghamSachsSolodukhin2002,ChenLongJHEP2010Cor,ChenLong2010,SetareKamali2010,GhezelbashKamali2010,ChenLongJHEP2010,ChenChen2011,ChenGhezelbash2011,Saktideformed2019,SaktiNucPhysB2020} by employing this duality that QNM computation is consistent with the gravity result, even for non-asymptotically flat solutions, for example for black holes with NUT charge and cosmological constant}. For the ECOs, it is crucial to note that the quantum effects that modify the near-horizon region can originate from the finite-size effects in CFT living on a torus with finite periodicities on its length $L$ and inverse of the temperature $1/T$. The finite-size effects on the CFT side are also believed to produce discreet QNMs spectrum. For ECOs, the QNM spectrum appears from the poles of the exponential parts of the CFT two-point function (see \ref{app:2point}). The two-point function is distinct from the CFT two-point function of the dual CFT for the black holes. The difference that appears is caused by the implementation of the method of images. By using this method, there will be an infinite sum in front of the two-point function. This infinite sum induces the emergence of the exponential parts. 

\textcolor{black}{In this paper, we basically compute QNM and other quantities from the CFT analysis. Moreover, for the ECOs with the exterior metric given by (\ref{KNNUTAdSmetric}) and from CFT analysis, the QNM is given by other poles of the exponential parts of the CFT two-point function. From Eq. (\ref{eq:forQNM}) and the frequencies given in Eq. (\ref{eq:omegageneral}), we obtain the following QNM spectrum
\begin{widetext}
\begin{equation}
\omega - \frac{m \Omega_H[r_+^2 +(a+n)^2]}{r_+^2 +r_*^2 +2(a+n)^2} +\frac{i\mathbf{a}}{2\pi L T_H}\frac{(\omega -m \Omega_H)a}{r_+^2 +r_*^2 +2(a+n)^2} =\frac{aR_n}{\left[r_+^2 +r_*^2 +2(a+n)^2\right]L}.\nonumber\
\end{equation}
\begin{equation}
\rightarrow \omega - m \Omega_H \simeq \frac{(2R_n+1+\delta)a}{2L\left[r_+^2 +r_*^2 +2(a+n)^2\right]} \left(1-i\mathbf{a}\frac{a\times \text{sign}(2R_n+1+\delta)}{2\pi L T_H(r_+^2 +r_*^2 +2(a+n)^2)} \right).\label{eq:QNMRastall}\
\end{equation}
\end{widetext}
The Eq. (\ref{eq:QNMRastall}) is obtained by expanding the factor containing a complex number to the first order.} Note that for electromagnetic perturbation, the factor $ (2R_n+1+\delta) $ becomes $ (2R_n+\delta) $. The QNM spectrum (\ref{eq:QNMRastall}) will reduce to the QNM spectrum of Kerr-like ECO \cite{DeyAfshordiPRD2020} when all parameters vanish, except $ M $ and $ a $. Interestingly, we can also find the QNM spectrums for various ECOs. Simply from QNM spectrum given in Eq. (\ref{eq:QNMRastall}), we can obtain the QNM of Kerr-Newman-NUT-like ECO when $ \alpha =0 $, QNM of Kerr-Newman-like ECO when $ n,\alpha =0 $, and QNM of Kerr-NUT-like ECO when $ e,g,\alpha =0 $.

\begin{figure*}[!hbtp]
	\centering
	\includegraphics[scale=0.85]{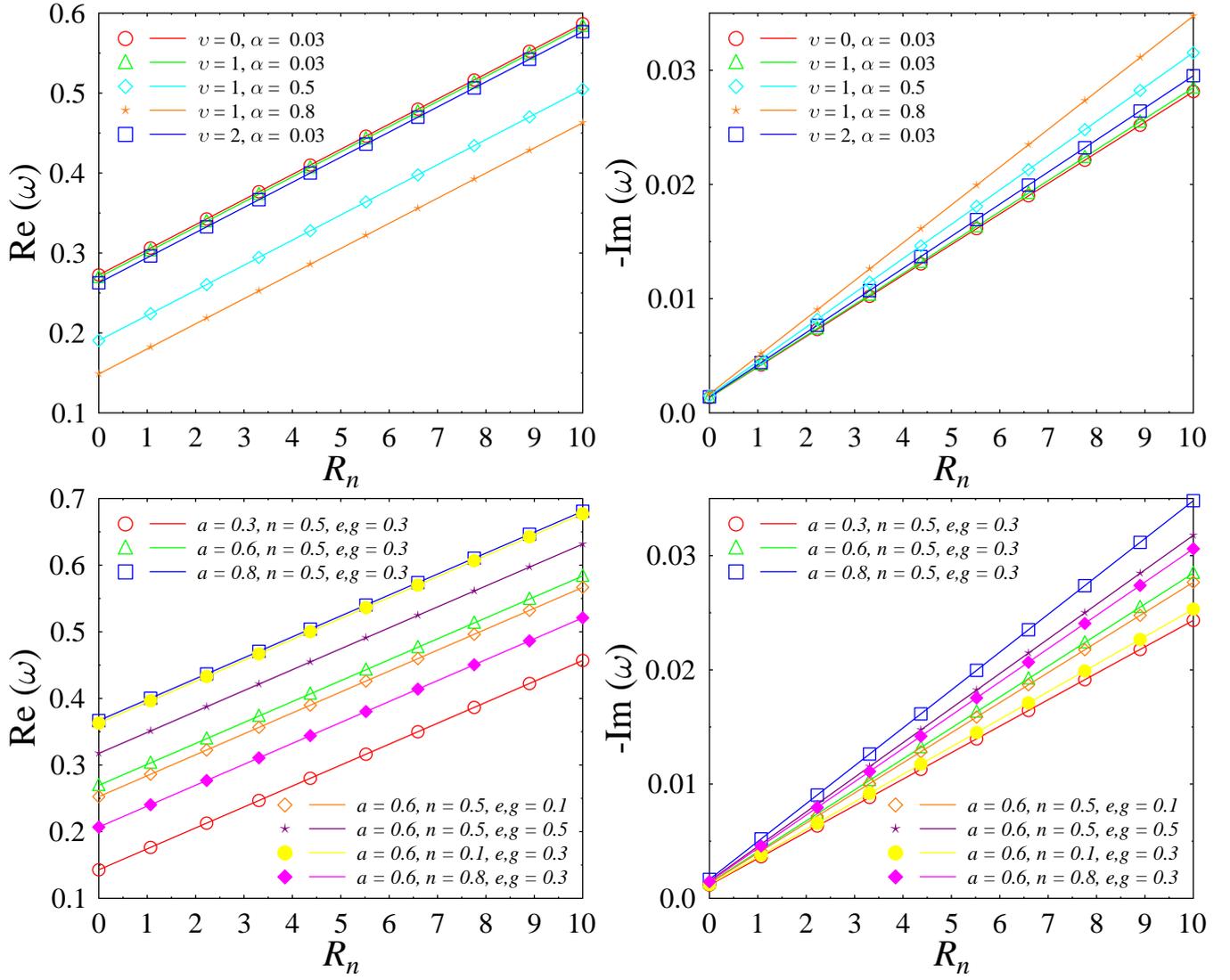}
	\caption{Real and imaginary parts of the QNMs as a function of integer $R_n$. with $v$ and $\alpha$ variations in upper panels and with $a$, $n$, and $e,g$ varitations in lower panels. We set $ M=1,r_0^*=100,m=2,\delta=0,\gamma=1 $ for all panels.}
	\label{fig:QNM}
\end{figure*}

The reflectivity of the near-horizon quantum membrane of the ECO can be obtained from the imaginary part of the QNM spectrum \cite{OshitaAfshordiPRDQuantum2020}. From the imaginary part of the QNM spectrum for Boltzmann ECOs (\ref{eq:QNMRastall}), we can find that the reflectivity of the membrane is given by
\begin{equation}
\mathcal{R}= e^{-\frac{|\omega -m\Omega_H|\mathbf{a}}{T_{H}}}, \label{eq:reflectivityCFT}\
\end{equation}
where we also need the modular parameter $ \mathbf{a} $ to be $ 1/2\gamma $ and the following relation for the length of the torus
\begin{equation}
L =\frac{a|r_0^*|}{\pi [r_+^2 +r_*^2 +2(a+n)^2]} \label{eq:lengthtorus}.\
\end{equation}
However, we can generalize the modular parameter $ \mathbf{a} $. Subsequently, it is important to see the role of every parameter to the QNM spectrum because it can be the plausible observational signatures of the quantum gravitational effects on this ECO. To see the role of the parameters, in Fig. \ref{fig:QNM}, we give the numerical value of the (real and imaginary) QNMs for different values of parameters to see the role of the parameters. From top panels, we can see that higher $\upsilon $ and higher $\alpha $ produce lower $\omega $. The bottom panels denote that higher $a, e $ and $g $ corresponds to higher $\omega $ while higher $ n $ corresponds to lower $\omega $. \textcolor{black}{It is worth noting that the information of $\omega_q$ and $\kappa\lambda$ analytically stored in the function $\upsilon$ as we can see in its definition after Eq. (\ref{KNNUTAdSmetric}). $\omega_q$ and $\kappa\lambda$ determine the number of the horizons (as explained in Ref. \cite{SaktiAnnPhys2020}). For several values of $\omega_q$ and $\kappa\lambda$, we can obtain a similar $\upsilon$. Hence, there will be a degeneracy on $\upsilon$, bot not on the values of $\omega_q$ and $\kappa\lambda$.}

\subsection{Absorption cross-section}\label{sec:PabsECO}

In the two-point function of the ECOs, the quantum membrane has contributed to produce new poles beside the poles coming from the gamma functions. In Eq. (\ref{eq:PabsECOgen}), it is seen that the presence of the quantum membrane very close to the would-be horizon gives an intriguing contribution to the absorption cross-section of the ECO. In addition, in the absorption cross-section, the presence of the quantum membrane contributes to the emergence of the echoes portrayed by the oscillatory feature. In the CFT description, the absorption cross-section of the ECO can be obtained by considering the dual field theory on a circle of length $ L $ and delve into the finite-size effects on the boundary. The detailed derivation of the absorption cross-section of the ECO is given in \cite{DeyAfshordiPRD2020}.

For the two-point function of 2D CFT, the absorption cross-section is given by the following expression 
\begin{eqnarray}
P^{\text{ECO}}_{\text{abs}} &\sim &\omega^{2l-1} {T _L}^{2h_L - 1} {T _R}^{2h_R - 1} \sinh \left( {\frac{{{\tilde{\omega}} _L }}{{2{T _L} }} + \frac{{{\tilde{\omega}} _R }}{{2{T _R} }}} \right)\nonumber\\
&\times & \left| {\Gamma \left( {h_L + i\frac{{{\tilde{\omega}} _L }}{{2\pi {T _L} }}} \right)} \right|^2 \left| {\Gamma \left( {h_R + i\frac{{{\tilde{\omega}} _R }}{{2\pi {T _R} }}} \right)} \right|^2. \nonumber\\
& \times & \frac{1-e^{-2\mathbf{a}\big|\frac{\tilde{\omega}_L}{T_L} +\frac{\tilde{\omega}_R}{T_R}\big|}}{\bigg|1-e^{2\pi i L(\tilde{\omega} _L+\tilde{\omega} _R)-\mathbf{a}\big|\frac{\tilde{\omega}_L}{T_L} +\frac{\tilde{\omega}_R}{T_R}\big|}\bigg|^2} \label{Pabs2CFTSL2ZECO}
\end{eqnarray}
The last factor is the source of the echoes where for the black holes, the last factor will vanish. The general left- and right-moving temperatures $ (T_L,T_R) $ are given by Eq. (\ref{eq:tempCFTKNNUT}). The CFT frequencies $ (\tilde{\omega}_L,\tilde{\omega}_R)$ are also identical with those of black holes which are given by Eq. (\ref{eq:omegageneral}). For our ECO, in the low-frequency limit, the absorption cross-section reads as
\begin{widetext}
	\begin{eqnarray}
	P^{\text{ECO}}_{\text{abs}} &\sim & \omega^{2h-1} \sinh \left( \frac{2\pi[r_+^2 +(a+n)^2]}{K(r_+ -r_*)}(\omega -m\Omega_H) \right)  \left| {\Gamma \left( h -\frac{i[r_+^2 +r_*^2+2(a+n)^2]\omega}{K(r_+ -r_*)}\omega +\frac{i2am}{K(r_+ -r_*)} \right)} \right|^2  \left| {\Gamma \left( h -\frac{i(r_+ +r_*)}{K}\omega \right)} \right|^2 \nonumber\\
	& & \times \frac{1-|\mathcal{R}|^2}{|1- \mathcal{R}e^{-2ir_0^*(\omega -m\Omega_H)+i\delta }|^2}. \nonumber\\
	\label{eq:abscrossRastall}\
	\end{eqnarray}
\end{widetext} Note that we have taken $ \mathbf{a}=1/2 $.

As the QNM spectrum, the absorption cross-section (\ref{eq:abscrossRastall}) will reduce to the absorption cross-section of Kerr-like ECO \cite{DeyAfshordiPRD2020} when only $ M $ and $ a $ which do not vanish and by changing $ h\rightarrow l $. It is worth to note that we can also find the absorption cross-section for various ECOs. From the absorption cross-section (\ref{eq:abscrossRastall}), we can obtain the absorption cross-section of Kerr-Newman-NUT-like ECO when $ \alpha =0 $, absorption cross-section of Kerr-Newman-like ECO when $ n,\alpha =0 $ and absorption cross-section of Kerr-NUT-like ECO when $ e,g,\alpha =0 $.

\begin{figure*}[hbtp]
	\centering
	\includegraphics[scale=0.75]{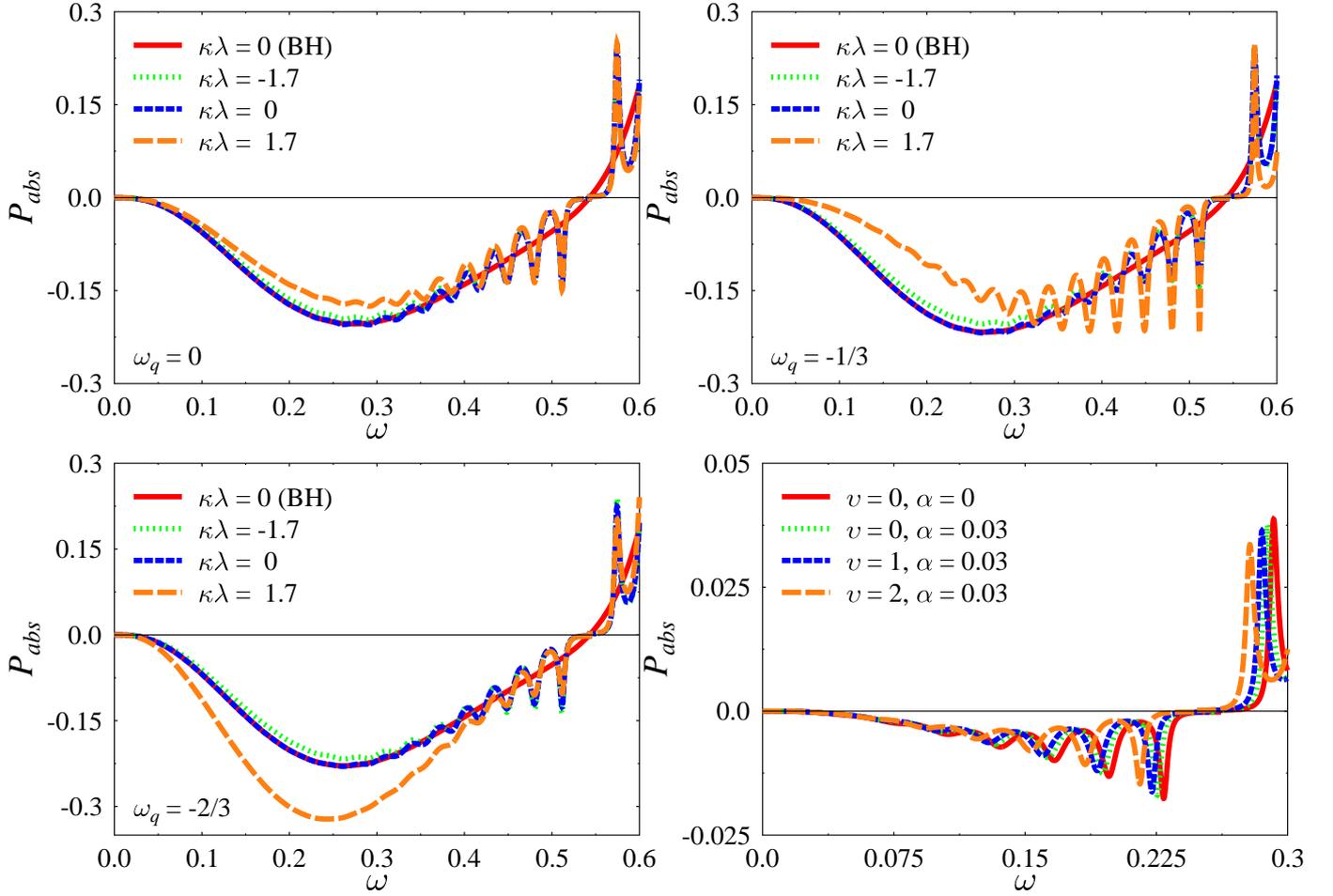}
	\caption{Absorption cross-section as a function of $\omega$ with variations of $v$, $\alpha$,$ \omega_q $, and $ \kappa\lambda $. We set $ m=2, C_s = -2, r_+ =1,  r_0^*=100,\delta=0, a=0.6, e,g=0.3, n=0.5 $ and $ \alpha=0.03 $.}
	\label{fig:Pabs1}
\end{figure*}

\begin{figure*}[hbtp]
	\centering
	\includegraphics[scale=0.75]{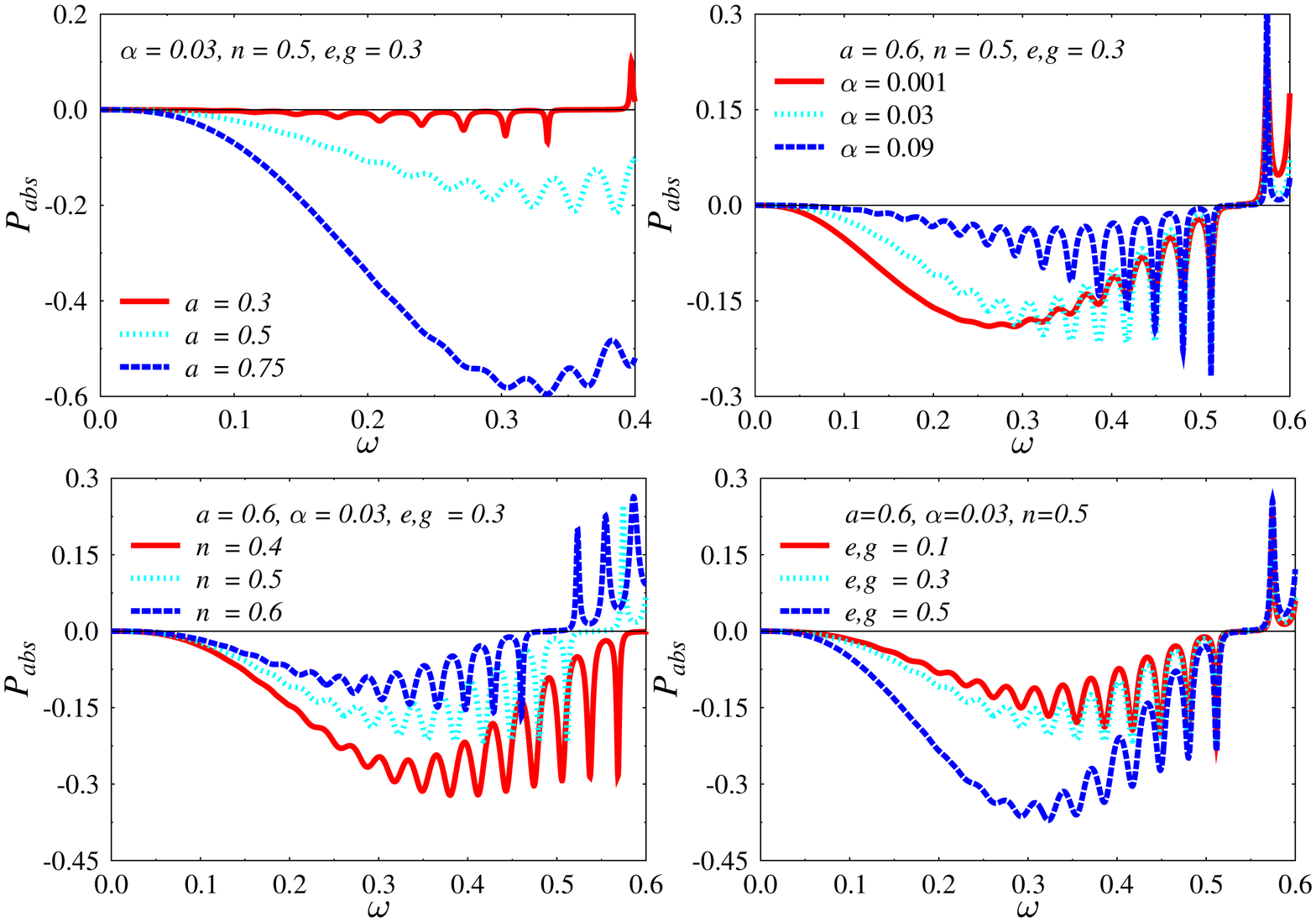}
	\caption{Absorption cross-section as the variation of $a, \alpha, n, e $ and $ g $. We set $\upsilon=1, m=2, l=2, r_0^*=100, \delta=0, M=1$.}
	\label{fig:Pabs2}
\end{figure*}

To show the membrane reflectivity's effect of the ECO and the parameters on the absorption cross-section, we need to plot the absorption cross-section by varying all parameters. We give the plot in Fig. \ref{fig:Pabs1}. The absorption cross-section in the low-frequency limit is negative due to superradiance. In the top panels and bottom left panel of Fig. \ref{fig:Pabs1}, we vary $ \kappa\lambda $ and $ \omega_q $ for the ECO and compare it to the black hole $ (\mathcal{R}=0) $. The absorption cross-section for the ECO possesses an oscillatory feature that corresponds with the reflectivity of the quantum membrane. For positive $ \kappa\lambda $, orange dashed line, the absorption cross-section is quite distinctive compared to vanishing and negative $ \kappa\lambda $ where for $\omega_q=-2/3$, the reflectivity is higher than those of when $\omega_q=0,-1/3$. The bottom right panel shows the plot of the variations of $ \upsilon $ and quintessential intensity $ \alpha $. The higher $ \upsilon $ results in the lower reflectivity. Moreover, we can see that for vanishing quintessential intensity, the reflectivity goes higher than the non-vanishing ones.  In Fig. \ref{fig:Pabs2}, we vary $a, \alpha, n, e $ and $ g $. From the top left panel, we can see that higher $ a $ makes $ P_{abs} $ more negative and has higher reflectivity. On the top right panel, it can be seen that higher $ \alpha $ corresponds to the lower $ P_{abs} $ and lower reflectivity. From the bottom panels, we can see that NUT charge $ n $ possesses similar feature as $ \alpha $ while $ e,q $ have similar feature as $ a $.

From Figs. \ref{fig:Pabs1} and \ref{fig:Pabs2}, we can also see the role of the parameters regarding the superradiant instability (also called ergoregion instability because it occurs near the ergoregion). For ECOs, the transmitted wave from the effective potential is reflected by the ECO's surface. This wave then passes through the ergoregion and extracts the rotational energy from the ECO (Penrose's process). A part of these waves which has gained energy from the ECO will have a higher amplitude and experience similar processes. Since these processes are repeating, the amplitude will grow exponentially, leading to instability. The instability always occurs when the real part of the frequency satisfies the superradiant condition $ \omega<m\Omega_H $. From the absorption cross-section analysis, this instability is depicted in the region below the superradiant bound. From Fig. \ref{fig:Pabs1}, we can analyze that $ \kappa\lambda $ affects the instability by increasing or reducing the absorption cross-section. For $ \omega_q=0, -1/3 $, the coupling constant may reduce the instability, while for $ \omega_q=-2/3 $, the instability increases for the positive coupling constant and decreases for the negative one. In Fig. \ref{fig:Pabs2}, we can see that the increasing value of the spin parameter and electromagnetic charges may increase the instability while the increasing value of quintessential intensity and NUT charge may reduce the instability.

\subsection{Echo time delay}\label{sec:echotime}

Since there is a reflective membrane near the would-be horizon, the ingoing wave is partially reflected by the membrane while copies of the reflected wave reach infinity after some time delay. This time delay is said to be the time for the wave to travel from the angular momentum barrier to the quantum membrane and travel back to the angular momentum barrier again. The time gap between two consecutive echoes can be determined in terms of the black hole parameters and the location of the reflective membrane. The echoes time delay can be computed as
\begin{equation}
\Delta \tau = 2|r_0^*|,\label{eq:echotimeawal}
\end{equation}
where $ r_0^* $ is the position of the reflective membrane in tortoise coordinate. The tortoise coordinate is given by $ r^*=\int [r^2+(a+n)^2]dr/\Delta $.
In the near-horizon region, the integral will become
\begin{equation}
r^*\simeq \int \frac{r^2+(a+n)^2}{K(r-r_+)(r-r_*)}dr.
\end{equation}
By integrating above equation in the near-horizon region and using (\ref{eq:echotimeawal}), we find the echo time delay as
\begin{equation}
\Delta \tau= \frac{2[r_+^2+(a+n)^2]}{K(r_+ - r_*)}\ln\left( \frac{\delta r_+}{r_+ - r_*} \right). \label{eq:echotime}\
\end{equation}
The constant $ \delta r_+ $ is very small that can be calculated from Eq. (\ref{eq:epsilon}). Hence, we can obtain
\begin{equation}
\delta r_+ = \frac{\eta^2 l_p^2 K(r_+ - r_*)}{4M[r_+^2 + (a+n)^2]} .\label{eq:epsilonKNUT}\
\end{equation}
From Eq. (\ref{eq:echotime}), we can see that the echo time delay is sensitive to the value of the position of the membrane in front of the horizon. Moreover, the membrane position is determined by the constant $ \eta $, which is related to the length of the torus $ L $ and proportional to the CFT degree of freedom $ N $. It is evident also that the echo time delay depends on all parameters, including the Rastall coupling constant. 
\begin{figure*}[hbtp]
	\centering
	\includegraphics[scale=0.75]{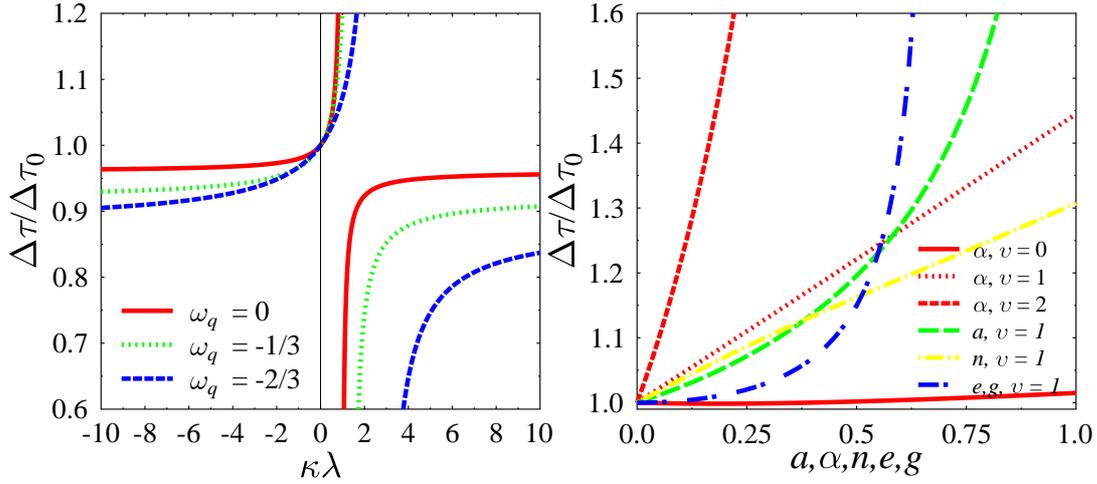}
	\caption{Echo time delay as the variation of $\kappa\lambda,\omega_q $ (left) and $ a,\alpha,n,e,g $ (right). For the left panel, we set $ r_+ = 1, a = 0.6, e,g = 0.3, n = 0.5; \alpha = 0.03,\eta^2 = 10^{30}$ and $ \Delta \tau_0 $ is the time delay for Einstein gravity. For the right panel, we set $ \upsilon =1 $ for the variation of $ a,n,e,g $ and $ M=1 $. $ \Delta \tau_0 $ is the time delay for the zero value of the parameters that we vary on the right panel. }
	\label{fig:echotime}
\end{figure*}

The black hole parameters, including the Rastall coupling constant, can modify the time delay between two successive echoes. Hence, by observing the time delay of the ringdown waveform of the ECOs, one can observe the parameters, especially to determine the Rastall coupling constant. When we can quantify the deviation from the Einstein gravity, we can expect that there exists gravitational theory beyond Einstein theory, which is, in this case, Rastall theory. In Fig. \ref{fig:echotime}, we show the plot of the echo time delay normalized by the time delay for vanishing parameters. For example, when we vary $ n $, it means that $ \Delta \tau_0 $ is the time delay when $ n=0 $. From the left panel, we can see that for every arbitrary $ \omega_q $, an irregular time delay corresponds to specific $ \kappa\lambda $. The irregularity occurs because $ \upsilon $ can be infinite for arbitrary $ \omega_q $ and $ \kappa\lambda $. Despite that, generally, we can observe for the ECOs in Rastall gravity that the presence of the Rastall coupling constant hastens the echoes compared to the ECO in general relativity, except when $\kappa\lambda $ approaches the irregularity point from the left side. The positive coupling constant hastens the echoes more than the negative one, except near the irregular point. In \cite{AbediDykaarPRD2017}, they note that the first echo in the time delay could be affected by nonlinear physics, so there can be a magnitude change on the time delay at $2\%$-$3\%$ level, and it might be detected in the future observations. In Rastall gravity, for example, we can find the change of magnitude at that level when  $\sim 0.32 \leq \kappa\lambda \leq 0.41 $ and $ \omega_q =0 $. These results provide fair information on the existence of a non-conserved energy-momentum tensor in a strong gravity regime that can be probed from echo time delay when we have the real observational data of echo time delay. We also argue that the quantum effects on the near horizon region can be manifested as a non-zero Rastall coupling constant and can be observed from the echo time delay.

Besides the Rastall coupling constant, other parameters can also modify the time delay leading to the delayed echoes. As we can see from the right panel of Fig. \ref{fig:echotime}, the increasing value of the parameters leads to the increasing of time delay, yet with different curves. The red solid, dotted, and dashed lines are the curves for the $ \alpha $ variation for different $ \upsilon $. For the same $ \alpha $, higher $ \upsilon $ will produce higher time delay. It is also seen that for higher $ a,n,e,g $, the produced time delay will be higher, denoting that the echoes will be delayed.

\section{Summary}\label{sec:summary}

In this work, we performed the Kerr/CFT correspondence computation to delve into the physical properties of rotating black holes and ECOs in Rastall theory of gravity in which the KNUTK black hole solutions in Rastall gravity have been chosen. Firstly, it has been shown that there exists a hidden conformal invariance in the low-frequency scattering off KNUTK black hole in Rastall gravity. Although the conformal symmetry is broken by periodic identification in the configuration space, it acts on the solution space of the scalar probe and associates a dual CFT description to the black hole. To be specific, the KNUTK black holes in Rastall gravity are dual to 2D CFTs with temperatures (\ref{eq:tempCFTKNNUT}) and central charges (\ref{centralchargeJpic}). Nevertheless, since we wanted to have regular central charges and non-vanishing temperatures, we required that $K\neq 0$ which corresponds to $ \alpha \upsilon (\upsilon-1)r_+^{\upsilon-2} \neq 2$. From the Cardy formula counting the microstate degeneracy in the dual CFT, we recovered the Bekenstein-Hawking entropy. For the massless neutral scalar scattering, we identified the dual operators with conformal dimension, left and right frequencies, which allow us to find the perfect match with the CFT prediction. Our findings on this black hole support that the conformal symmetry in the solution space of the scalar probe is essential to set up a CFT description. Henceforth, not only on the extremal condition \cite{SaktiPhysDarkU2020}, yet the non-extremal rotating black holes in Rastall gravity are also holographically dual with CFTs.

We also analyzed the holographic CFT description of an ECO for which the exterior space-time resembles the KNUTK black hole in Rastall gravity. There exists a reflective quantum membrane very close to the would-be horizon of the ECO, which produces distinct physical properties, especially the oscillatory feature on the absorption cross-section compared to those of black holes. We have implemented the dual CFT description of this ECO to compute the QNM spectrum where the spectrum comes from the poles on the exponential parts of the two-point function in the upper and lower half of $\omega$-plane. The QNM spectrum is very sensitive to the change of the black hole's parameters, including the Rastall coupling constant, resulting in several consequences. For example, higher $\upsilon $ produces a lower QNM spectrum. 

From the calculation of the absorption cross-section, the Rastall coupling constant and other parameters possess significant contributions to the superradiant instability of the ECO where the sign of the Rastall coupling constant is notably essential to increase or decrease the instability. We also computed the echo time  delay as the necessary observable from the gravitational echoes observation in the postmerger ringdown signal. Besides the quantum gravitational effects that could modify the near-horizon region, it could also be manifested as a non-conserved matter tensor giving rise to a non-vanishing Rastall coupling constant. It was shown that the Rastall coupling constant played a significant role in time delay. For instance, near the irregular point from the negative sign side of the coupling constant, the time delay gets higher, which denotes the change of magnitude in time delay coming from the correction of non-linear physics.  In a sense, the non-vanishing Rastall coupling constant could be grasped as the phenomenological realization of that correction. \textcolor{black}{Black hole and ECO considered in this paper contain many parameters which are $M, a, e, g, n, \alpha, \omega_q$, and $ \kappa\lambda$. It is very important to observe the effect of each parameter. We believe that from our calculation on the QNM spectrum, absorption cross-section and time delay, the effects of each parameter can be studied effectively. Hence, our analysis within this paper tries to provide the possibility of probing Rastall gravity in the black hole and ECO}. We believe that the future observations from interferometric detectors at higher sensitivity will confirm these findings of which the time delay can probe the existence of the Rastall gravity.

\section*{Acknowledgments}
M. F. A. R. Sakti is partially supported by C2F Fellowship, Chulalongkorn University. A. Sulaksono is partially supported by DRPM UI's (Skema PPI Q1 2021) Grants No. NKB-586/UN2.RST/HKP.05.00/2021. A. Suroso and F. P. Zen are supported in part by Riset ITB. F. P. Zen is also supported by the Ministry of Education, Culture, Research and Technology of Indonesia.


\appendix
\section{CFT Two-point Function}
\label{app:2point}
It has been conjectured that the quantum gravity living in the near-horizon region of the Kerr-like ECO is dual to a 2D CFT living on a circle of a finite length $ L $ \cite{DeyAfshordiPRD2020}. Similarly with those in the  black holes, for ECOs, the CFT coordinates are defined by 
\begin{equation}
t_L= -2\pi T_L \phi -2n_L t, ~~~ t_R= 2\pi T_R \phi +2n_R t.
\end{equation}
We can also apply the following azimuthal periodicity with the addition of thermal periodicity of imaginary time
\begin{equation}
\phi \rightarrow \phi +2\pi L + i\frac{\Omega_H}{T_H},~~~ t= t +\frac{i}{T_H}.
\end{equation}
Under the above identifications, the CFT coordinates now become
\begin{eqnarray}
(t_L,t_R) &\sim & (t_L,t_R) + \left(-4\pi^2 T_L L -2\pi i T_L \frac{\Omega_H}{T_H} \right. \nonumber\\
& & \left. -2\pi i \frac{n_L}{T_H}, 4\pi^2 T_R L + 2\pi i T_R \frac{\Omega_H}{T_H} \right).\
\end{eqnarray}

The QNM spectrum can be found from the poles of the retarded CFT correlation function. However, the correlation function that will be used is generalized for general value of modular parameters ($ \mathbf{a,b,c,d} $). It is worth performing Fourier transformation of the retarded CFT correlation function to the momentum space. The Fourier transform of the two-point function for the general value of modular parameters is given as
\begin{eqnarray}
\tilde{G}(\omega_L,\omega_R) &=& \int dt_L dt_R e^{-i\omega_L T_L}e^{-i\omega_R T_R}\nonumber\\
&\times& \sum_{\bar{p}\in\mathcal{Z}}\frac{(\pi T_L)^{2h_L}}{\left[\sinh\left\{\pi T_L\left(\frac{t_L}{2\pi T_L}+2\pi \bar{p}L+\frac{i\mathbf{a}\bar{p}}{T_L}\right)\right\}\right]^{2h_L}}\nonumber\\
&\times& \frac{(\pi T_R)^{2h_R}}{ \left[\sinh\left\{\pi T_R\left(\frac{t_R}{2\pi T_R}+2\pi \bar{p}L+\frac{i\mathbf{a}\bar{p}}{T_R}\right)\right\}\right]^{2h_R}}.\
\end{eqnarray}
Then by defining the following new coordinates
\begin{eqnarray}
&&\tilde{t}_R = \frac{t_R}{2\pi T_R}+2\pi \bar{p}L+\frac{i\mathbf{a}\bar{p}}{T_R}, \nonumber\\
&& \tilde{t}_L =\frac{t_L}{2\pi T_L}+2\pi \bar{p}L+\frac{i\mathbf{a}\bar{p}}{T_L},\
\end{eqnarray}
one can find

\begin{widetext}
	\begin{eqnarray}
	\tilde{G}(\tilde{\omega}_L,\tilde{\omega}_R)&\sim & \sum_{\bar{p}\in\mathcal{Z}}e^{i\bar{p}\left(2\pi \tilde{\omega}_L L + 2\pi \tilde{\omega}_R L +i\mathbf{a}\frac{\tilde{\omega}_L}{T_L}+i\mathbf{a}\frac{\tilde{\omega}_R}{T_R}\right)} \int  \frac{d\tilde{t}_L d\tilde{t}_R e^{-i\tilde{\omega}_L T_L -i\tilde{\omega}_R T_R} \left(\pi T_L \right)^{2h_L}\left(\pi T_R \right)^{2h_R}}{\left[\sinh(\pi T_L \tilde{t}_L)\right]^{2h_L}\left[\sinh(\pi T_R \tilde{t}_R)\right]^{2h_R}}\nonumber\\
	& \propto & {T _L}^{2h_L - 1} {T _R}^{2h_R - 1} e^{-{\frac{{\tilde{\omega} _L }}{2{T _L} } - \frac{{\tilde{\omega} _R }}{2{T _R} }}}\left| {\Gamma \left( {h_L + i\frac{{\tilde{\omega} _L }}{2\pi {T _L} }} \right)}  {\Gamma \left( {h_R + i\frac{{\tilde{\omega} _R }}{2\pi {T _R} }} \right)} \right|^2 \left[\frac{1}{1-e^{2\pi i L(\tilde{\omega} _L+\tilde{\omega} _R)-\mathbf{a}\big|\frac{\tilde{\omega}_L}{T_L} +\frac{\tilde{\omega}_R}{T_R}\big|}} -\frac{1}{1-e^{2\pi i L(\tilde{\omega} _L+\tilde{\omega} _R)+\mathbf{a}\big|\frac{\tilde{\omega}_L}{T_L} +\frac{\tilde{\omega}_R}{T_R}\big|}} \right] . \nonumber\\
	\label{eq:greenfunctECO}\
	\end{eqnarray}
\end{widetext}
In above two-point function, there are two types of poles emerging from the gamma functions and exponential parts lying in the upper and lower half of the $ \omega $-plane. It has been shown in \cite{ChenChuJHEP2010,ChenLongJHEP2010Cor} that the QNM from the poles emerging from the gamma function agree with the QNM spectrum of the black holes. In addition, the poles emerging from the exponential parts will result in the QNM spectrum for ECOs. In order to produce the QNM spectrum for ECOs, we choose the pole in the lower half plane which gives
\begin{equation}2\pi i L(\tilde{\omega} _L+\tilde{\omega} _R)-\mathbf{a}\big|\frac{\tilde{\omega}_L}{T_L} +\frac{\tilde{\omega}_R}{T_R}\big|=2\pi i R_n, \label{eq:forQNM}
\end{equation}
where $ R_n $ is positive integer. From above equation, we can approximate the value of QNM  spectrum.



 \bibliographystyle{elsarticle-harv}

\end{document}